\RequirePackage[l2tabu, orthodox]{nag}
\documentclass[envcountsect,10pt]{llncs}
\usepackage[charter]{mathdesign}

\usepackage[T1]{fontenc}
\usepackage[utf8]{inputenc}
\usepackage{textcomp} % Euro and other unicode symbols
\PassOptionsToPackage{hyphens}{url}
\usepackage[breaklinks,colorlinks=true]{hyperref}
\usepackage{xcolor}
\usepackage{graphicx}
\usepackage{overpic}
\usepackage{amsmath}
\usepackage[english]{babel}
\usepackage[style=numeric,maxnames=5,giveninits=true,doi=false,isbn=false,url=false,eventdate=comp]{biblatex}
\usepackage[extralong]{bibstrings}
\usepackage{xspace}
\usepackage{verbatim}
\usepackage{general}
\usepackage{markup}
\usepackage{annot}
\usepackage{bodyversion}
\usepackage{environment}
\usepackage{hyphenation}

\AtEveryBibitem{%
\ifentrytype{inproceedings}{
  \clearname{editor}%
  \clearfield{booktitleaddon}%
}{}
}

\def\setoverpicfontsize{\relax}

\newcommand{\user}{U}
\newcommand{\useraddr}{\hat{U}}
\newcommand{\issuer}{I}
\newcommand{\attr}[1][{}]{a_{#1}}
\newcommand{\attrs}{\attr[1],\ldots,\attr[m]}
\newcommand{\cred}[2]{C_{#2}(#1)}
\newcommand{\credentialdetails}{\cred{\attrs}{\PTO}}
\newcommand{\sig}[2]{[#1]_{#2}}
\newcommand{\grey}[1]{%
  \begingroup
  \setlength{\fboxsep}{0.5pt}%
  \colorbox{lightgray}{\rule[-1.5pt]{0pt}{8.5pt}$#1$}%
  \endgroup
}
\newcommand{\blindsig}[3]{\sig{\grey{#1}\,|\,#2}{#3}}
\newcommand{\useblindsig}[3]{\sig{#1\,|\,#2}{#3}}
\newcommand{\trip}[1][]{T_{#1}}
\newcommand{\tripdetails}[1][{}]%
  {\langle \triproute[#1],\tripdate[#1]\rangle}
\newcommand{\triproute}[1][]{r_{#1}}
\newcommand{\tripdate}[1][]{d_{#1}}
\newcommand{\fare}[1][]{{f_{#1}}}
\newcommand{\farefor}[1]{\textit{fare}(#1)}
\newcommand{\PSP}{\textrm{PSP}}

\newcommand{\receiptseqno}[1][]{{\textit{rs}_{#1}}}
\newcommand{\receipt}[1][]{{R_{#1}}}
\newcommand{\receiptdetails}[1][]{%
  \blindsig{\receiptseqno[#1]}{\fare[#1]}{\privkey{\PSP}}}
\newcommand{\usereceiptdetails}[1][]{%
  \useblindsig{\receiptseqno[#1]}{\fare[#1]}{\privkey{\PSP}}}
\newcommand{\PTO}{\textrm{PTO}}
\newcommand{\ticket}[1][]{T_{#1}}
\newcommand{\ticketseqno}[1][]{\textit{ts}_{#1}}
\newcommand{\ticketdetails}[1][]{%
  \blindsig{\trip[#1],\ticketseqno[#1]}{\fare[#1]}{\privkey{\PTO}}}
\newcommand{\useticketdetails}[1][]{%
  \useblindsig{\trip[#1],\ticketseqno[#1]}{\fare[#1]}{\privkey{\PTO}}}
\newcommand{\PTC}{\textrm{PTC}}
\newcommand{\creditvalue}[1][]{v_{#1}}
\newcommand{\creditseqno}[1][]{\textit{cs}_{#1}}
\newcommand{\credit}[1][]{C_{#1}}
\newcommand{\creditdetails}[1][]{%
  \blindsig{\creditseqno[#1]}{\creditvalue[#1]}{\privkey{\PTC}}}
\newcommand{\usecreditdetails}[1][]{%
  \useblindsig{\creditseqno[#1]}{\creditvalue[#1]}{\privkey{\PTC}}}
\newcommand{\creditreceiptdetails}[1][]{%
  \blindsig{\receiptseqno[#1]}{\creditvalue[#1]}{\privkey{\PSP}}}
\newcommand{\usecreditreceiptdetails}[1][]{%
  \useblindsig{\receiptseqno[#1]}{\creditvalue[#1]}{\privkey{\PSP}}}
\newcommand{\checkin}[1][]{\textrm{I}_{#1}}
\newcommand{\checkindetails}[1][]{%
  \sig{\creditseqno[i],\location[#1],\ctime[#1]}{\privkey{\PTO}}}
\newcommand{\usecheckindetails}[1][]{%
  \sig{\creditseqno[i],\location[#1],\ctime[#1]}{\privkey{\PTO}}}
\newcommand{\location}[1][]{\ell_{#1}}
\newcommand{\ctime}[1][]{t_{#1}}
\def\setoverpicfontsize{\scriptsize}
\title{Privacy Friendly E-Ticketing For Public Transport\thanks{%
  Version: Fri Jan 22 10:58:06 2021 +0100 / arXiv / ov-pet.tex}
}
\author{Jaap-Henk Hoepman\inst{1,2}}
\institute{
  Radboud University Nijmegen, Email: \email{jhh@cs.ru.nl} \and
  University of Groningen
}
\begin{document}
\maketitle
\begin{abstract}{
  This paper studies how to implement a privacy friendly form of ticketing for public transport in practice. The protocols described are inspired by current (privacy invasive) public transport ticketing systems used around the world. The first protocol emulates paper based tickets. The second protocol implements a pay-as-you-go approach, with fares determined when users check-in and check-out. Both protocols assume the use of a smart phone as the main user device to store tickets or travel credit. We see this research as a step towards investigating how to design commonly used infrastructure in a privacy friendly manner in practice, paying particular attention to how to deal with failures.}
\end{abstract}

\section{Introduction}

At the turn of the century, several countries transitioned from paper based tickets for public transport to electronic forms of ticketing, either for public transport in a metropolitan area like the London Underground (the so-called Oyster card\footnote{%
  \url{https://oyster.tfl.gov.uk/}
})
and Hong Kong public transport (the Octopus Card\footnote{%
  \url{https://www.octopus.com.hk/en}
}),
or for all public transport in an entiry country (like the OV-chipkaart\footnote{%
\url{https://www.ov-chipkaart.nl/}
}
in the Netherlands). These e-ticketing systems are typically based on contactless smart cards. Some of these systems exhibit significant weaknesses in terms of security~\autocite{DBLP:conf/esorics/GarciaGMRVSJ08} (because the smart cards used need to be cheap and therefore contain weaker security features) and privacy~\autocite{gudymenko2015eticketing} (because these smart cards often contain a unique fixed identifier~\autocite{garfinkel2005rfid-privacy,juels2006rfid-secpriv-survey}).\footnote{%
  Note that these security weaknesses are not necessarily easily exploitable in practice~\autocite{tno2008ov-cipkaart}.
}
%\autocite{deKGGH2008}

In this paper we focus on the privacy issues in e-ticketing for public transport. This problem has been studied in the past for e-ticketing in particular~\autocite{DBLP:conf/pet/Heydt-BenjaminCDF06,hinterwalder2013efficient-ecash}, but also for related problems like electronic toll collection systems~\cite{fetzer2020p4tc} or more general road pricing systems~\autocite{hoepman2010roadpricing}. Compared to the work of Heydt-Benjamin~\etal~\autocite{DBLP:conf/pet/Heydt-BenjaminCDF06,hinterwalder2013efficient-ecash} (which relies on anonymous e-cash and anonymous credentials as building blocks) we use (partially) blind signatures instead to create either unlinkable travel tickets or unlinkable travel credit. This makes the protocols less complex and more efficient as there is no need to spend several e-cash coins to pay the exact fare. Moreover, our approach is inspired by the work of Stubblebine~\etal~\autocite{stubblebine1999unlinkable-transactions}, which studied unlikable transactions in practice, with a particular focus on dealing with failures.

We present two privacy friendly e-ticketing protocols for public transport. The first protocol emulates paper based tickets. The second protocol implements a pay-as-you-go approach, with fares determined when users check-in and check-out. Both protocols assume the use of a smart phone as the main user device to store tickets or travel credit, without relying on any tamper-proof component. We wish to stress that this means that we put no trust assumptions on the device the user uses to pay for public transport. In both cases we pay particular attention to possible failures and  how to graciously deal with them. 

The remainder of the paper is structured as follows. We introduce the system model, requirements, threat model and other assumptions in section~\ref{sec-problem}. We discuss protecting privacy in practice in section~\ref{sec-practice}, especially the assumption to use smartphones as the primary user device. Section~\ref{sec-primitives} discusses the primitives used in our protocols, that are then presented in section \ref{sec-papertickets} and section \ref{sec-pay-as-you-go}. Our conclusions are presented in section~\ref{sec-concl}.
% Related work in \label{sec-related}.

\section{Problem statement}
\label{sec-problem}

%\subsection{Stakeholders and system model}

We assume a system that supports many modes of transportation. This means we distinguish several \emph{public transport operators (PTOs)} that offer public transport services. \emph{Users (U)} travelling by public transport make trips that may consist of several legs each using a different mode of transportation offered by a different PTO. \emph{Inspectors} on the trains and busses verify that everybody on board has a valid ticket. A central \emph{public transport clearinghouse (PTC)} provides the public transport ticketing infrastructure (or at least the APIs to connect to this infrastructure), and distributes financial compensation to the PTOs for services rendered. Separate \emph{payment service providers (PSPs)} handle payments from banks initiated by users.

Users have a digital \emph{token} that enables them to travel by public transport. Instead of relying on a smart card (to store tickets or other information needed to verify whether someone is entitled to a certain mode of public transportation) we assume most people own a sufficiently modern and capable smartphone, and are willing to use it for public transport (we will discuss this further in section~\ref{ssec-threatmodel}). Any entity (in particular any of the PTOs) can offer an app for this purpose.

\subsection{Requirements}

Any public transport ticketing scheme for the model outlined above should satisfy the following requirements.

\begin{itemize}
\item
  Users should pay for trips, where the fare depends on when and where a user travels, the distance travelled, and whether the user has subscribed to a public transport pass that offers reduced fares (during certain times of the day, or on certain tracks).
\item
  Public transport operators should receive compensation for their services, which (partly) depends on all the actual trips made bu users that traversed part of their infrastructure. In other words, the amount of compensation can depend on how many passengers travelled on which particular track on which day.
\item
  The scheme should be privacy friendly: no party should be able to link any number of trips to each other (as belonging to one, unknown, person) or to any one particular person. In other words, the previous requirement only allows the PTO to learn \emph{how many} passengers travelled a certain track at a certain time of the day, not \emph{who} they were, or whether they were the \emph{same people} that set out on some other trip earlier.
\item
  The scheme should be secure: it should prevent or detect fraud by users (e.g. creating fake tickets, paying less than the required fare) and prevent fraud by operators (e.g. claiming more trips than actually took place over their infrastructure). Travelling without a valid `ticket' should be discouraged by regular inspection and appropriate fines.
\item
  The scheme should be fast enough to process large volumes of travellers at peak hours. Checking travellers by conductors should be fast, \eg take less than a second. Checking in or out to enter or exit public transport (like in the London Oystercard system or the Dutch OV-chipkaart system) should take only a few hundred milliseconds at most.
\end{itemize}
We note that the timing constraints mentioned in the last requirement are important in practice, but performance measurements are unfortunately out of scope for this paper.

\subsection{Threat model}
\label{ssec-threatmodel}

We assume users may try to actively defraud the system (travelling by public transport while paying less than required or nothing at all), if the probability of being caught is low. They will
%buy fraudulent smart cards or
root their smartphones and install fraudulent apps if there is a clear benefit. This means that in terms of smartphones we cannot assume any trusted environment to store secrets (we rule out the possibility that a public transport app gets to use a secure enclave on the smartphone). In other words: the device and the app are untrusted from the PTO perspective. This means our system is weaker than one relying on smart cards that \emph{can} be used to store secrets and keep them confidential, preventing their users from accessing (and perhaps copying) them.

We assume banks, PSPs, PTOs and the PTC will actively (and collectively) try to break privacy and recover trip details from their users, using any information they can get their hands on. They are untrusted from the user perspective.

We do assume however that PTOs do \emph{not} try to break privacy by writing their apps in such a way that the information provided by the user through the app, but shielded from the central PTO servers by the protocol, is surreptitiously sent to the PTOs regardless. The PTOs could in theory do this. We can mitigate this by offering third party (open source) apps, requiring external audits and analysis, or through the vetting procedures enforced by the smartphone app stores. (This, by the way, is another reason why we cannot assume that the smartphone or app can store secrets.)

We assume that PTOs will try to defraud the system and claim more compensation from the PTC than warranted. The PTC is trusted, in the sense that it does not favour one PTO over the other, and that at the of the day all money received must be spent (on compensating PTOs or on the cost of running the clearinghouse and its ticketing infrastructure). Audits can be used to ensure this.

We assume the cryptographic primitives used cannot be broken, and that entities keep their secrets secret (unless they could benefit from not doing so).

\subsection{Other assumptions}

We assume secure, i.e. authenticated and encrypted, connections between all entities. Clearly the user is not authenticated.
We assume fares are course enough to ensure that the price associated with a trip does not reveal the actual trip itself. For example, trip prices could be  set at fixed amounts for every ten kilometres travelled, with a fixed ceiling fare for all trips longer than a certain distance. (Care should be taken to ensure that for every possible fare the number of different trips with that fare is sufficiently high to guarantee a reasonable degree of anonymity.)

We also assume that local device to device communication is using only ephemeral identifiers (if any) to prevent linking devices over longer periods of time. This means WiFi or Bluetooth are using properly randomised MAC addresses, or random anti-collision identifiers if NFC is used. This also implies that we assume apps do not have access to any other permanent, unique, device specific identifier.\footnote{%
  The operating systems of these smartphones should, could and sometimes actually do prevent apps from having access to such a persistent identifier. Preventing the app itself to generate such an identifier itself and store it locally is of course not possible (although audits may reveal this).
} 

For normal (long range) internet connections between the smartphone and the servers of the other entities we cannot make such an assumption: it is rather trivial to track users based on their often fixed IP addresses. We discuss this in the next section.

\section{Protecting privacy in practice}
\label{sec-practice}

Protecting privacy in practice is a major challenge, for several reasons. First of all, practical considerations may rule out certain solutions or may make it impossible to make simplifying assumptions. For example a complex tariff system may lead to a situation where particular fares correspond to one or perhaps only a few particular trips. This is not the case when the tariff system is very simple (\eg two or three different zones in a metro network).

This issue is exacerbated when people are forced to pay for individual trips separately (see the first protocol that emulates paper tickets in section~\ref{sec-papertickets}) while the payment protocol is not anonymous. Except for cash payments (and certain privacy friendly crypto currencies perhaps) existing and widely accepted payment methods (credit card, debit card or e-banking apps) are account based and thus identifying.

Even if this were not the case, the protocols detailed below rely extensively on Internet connectivity that by its very nature is identifying. Strategies to shield the user's IP address from the other parties involved in the ticketing system (like using Tor~\autocite{dingledine2004tor} or mix networks~\autocite{chaum1981untraceable-mail})) should be used, but are probably impractical to use extensively and reliably at scale.
% \Question{Any research on this?}
Then again, letting users use a trusted VPN would solve most of the problems as this would hide all users behind the IP address of the VPN provider. We will have something more to say about this later on. 

The biggest paradox, from a privacy perspective, is of course the use of a smartphone as the basic user device for buying, storing and using tickets. On the one hand it is an entirely personal device, capable enough to orchestrate the interactions with all other parties using complex privacy friendly protocols, with the possibility of a nice user friendly interface to boot. Moreover, people expect their smartphone to support their day to day activities, like paying in shops, these days. This makes a smartphone the natural, in fact unavoidable, choice as the user device.\footnote{%
  Although a fall back option should always be available those people that cannot afford to own a (recently modern) smartphone.
}
But clearly the use of smartphones comes with severe privacy risks. By design, mobile phone operators know the approximate location of all their subscribers (and can zoom in using a process called triangulation). With GPS, standard on smartphones, location is also readily available to the phone itself as well as all apps that were granted permission to location services. With the increasing complexity of smartphones and the huge app ecosystem, users have very little reason to trust their smartphone or to expect it to protect their privacy.

With these caveats in mind we follow a pragmatic approach in this paper, aiming for a strong enough technical protection of privacy under reasonable assumptions. No coalition of PSPs, PTOs and the PTC can link trips\footnote{%
  Either as bought in the protocol that emulates paper based tickets, or as implied by check-in and corresponding check-out events
}
to users, beyond what can be ascertained by observing the financial transactions of the users, knowledge of the tariff structure, and (partial) apriori knowledge of the travel patterns of a subset of these users. We do not solely rely on technical mechanisms however, but also depend on legal, societal and market incentives to keep the different stakeholders in check. All measures combined should ensure that the cost of obtaining privacy sensitive information in general outweighs the (business) benefit.

The tacit assumption in this work is that it is much safer, from a privacy perspective, to collect personal data locally on the user device, instead of centrally on the servers of the service providers. Clearly a malicious public transport app can collect and upload all this personal data surreptitiously. The assumption %(see above)
is that this cannot or will not happen. 

\section{Primitives}
\label{sec-primitives}

Our protocols for privacy friendly ticketing for public transport are based on three primitives, that we will describe in this section: partially blind signatures, attribute based credentials, and a mechanism to implement a form of privacy friendly payment with receipt.

\subsection{Partially blind signatures}
\label{ssec-pbs}

Blind signatures were introduced by David Chaum almost four decades ago \autocite{chaum1982blind-signatures}, as the fundamental building block to
implement a form of untraceable digital cash. His proposal was to
represent each digital coin as a unique serial number blindly signed by
the issuing bank. The unique serial number embedded in the coin would
prevent double spending, while the blind signature over the coin would
guarantee both \emph{untraceability} (by not knowing which coin was
signed) and \emph{unforgeability} (by signing the coins in the first
place).

In the protocols below we use a generalisation of this idea called partially blind signatures, introduced by Abe and Fujisaki~\autocite{abe1996partially-blind} and further investigated and optimised by Abe and Okamoto~\autocite{abe2000partial-blind-sig,okamoto2006partial-blind-sig}. In a partially blind signature scheme the messages to be signed consists of a secret part (only known to the user) and a public part (known to both the user and the signer). Issuing a blind signature involves an interactive protocol between the user and the signer, where the user blinds the secret in order to hide it from the signer.

In the protocols below we use these partially blind signatures to issue receipts and/or tickets where the receipt number or the trip details are kept secret. Because such receipts and tickets are only used once (in fact, we need to enforce that they are not used more than once), using simple signatures instead of full blown attribute based credentials (to be discussed further on) suffices. When describing our protocols we write $\blindsig{\id{secret}}{\id{public}}{\privkey{}}$ when issuing a blind signature over secret part $\id{secret}$ and public part $\id{public}$ (using private key $\privkey{}$ of the signer to sign it), and write $\useblindsig{\id{secret}}{\id{public}}{\privkey{}}$ when subsequently using it (revealing both the secret and the public part).

\subsubsection{Two faces of blindness}

Chaum explained blind signatures intuitively by showing how a blind
signature could be implemented in a traditional, non digital, setting
using carbon paper inside paper envelopes. To obtain a blind signature
on a secret message, a user could send the message inside a sealed
envelope to the signer, with the inside of the envelope covered with
carbon paper. The carbon paper ensures that if the signer signs the
envelope from the outside, the carbon paper transfers this signature to
the secret message inside the envelope. When the signer returns the
still sealed envelope (proving it didn't see the message) all the user
needs to do is to open the envelope to obtain the blindly signed
message.

This intuitive explanation clearly shows that the message stays hidden
from the signer. But this by itself is not enough to prevent a bank from
tracing a digital coin signed this way, even if it prevents the bank
from learning its serial number. In fact, if the bank signs each
envelope in a slightly different way, and remembers which way of signing
it used to sign each envelope, it can link actual signatures on messages
to the particular envelope on which he put the exact same signature.

In other words, in order to guarantee untraceability (sometimes also called \emph{unlinkability}), blind signatures need to guarantee two separate blindness properties:
\begin{description}
\item[message hiding]
The message to be signed is hidden from the signer.
\item[signature unlinkability]
Given a final blind signature on a message, the signer cannot determine
when it generated that particular signature.
\end{description}
To see that these are indeed different properties, observe that a scheme where signing the cryptographic hash of message $m$ (without revealing $m$ itself to the signer) is message hiding but clearly not unlinkable. In the protocols below we rely on both these properties to hold. Most (partially) blind signature schemes in fact satisfy both of them. This is in particular the case\footnote{%
  The (partial) blindness property is defined using a game where two messages $m_0$ and $m_1$ are randomly assigned to two users (based on a random bit $b$). Each user then requests a blind signature on its message from the signer. The signer is then given both signatures (and for each the corresponding message) and asked to guess the value of $b$. If it could distinguish which signature corresponds to which user, it could for sure determine the value of $b$.
}
for the schemes of Abe and Okamoto~\autocite{abe2000partial-blind-sig,okamoto2006partial-blind-sig} (but not for the blind signature scheme underlying the Idemix attribute based credential scheme~\autocite{idemix2012,camenisch2001anoncred}).

\subsubsection{Dealing with failures}
\label{ssec-failures}

In the protocols below, partially blind signatures are used to represent receipts received after a successful payment, or as public transport tickets received in exchange for a valid receipt. In both cases a kind of `fair exchange'~\autocite{pagnia2003fair-exchange} is required between a user and a signer, and there should be a way to recover from errors in case messages are dropped, connections fail, or system components crash, to ensure that either the exchange takes place completely, or that the exchange is cancelled and both parties return to the state before they started the exchange. 

Recall from section~\ref{ssec-threatmodel} that we assume users to be malicious while service providers (the signers in this case) are honest (but curious). This assumption makes it possible and relatively easy to implement a fair exchange in this particular case. Details will vary depending on the particular blind signature scheme used.

For example, the partial blind signature scheme of Okamoto~\autocite{okamoto2006partial-blind-sig} consists of the following phases when creating the blind signature $\blindsig{s}{p}{\privkey{}}$.
\begin{enumerate}
\item
  \label{item-blind}
  The user blinds the secret part $s$ using some randomness $r_u$ as
  $b =\id{blind}(s,r_u)$ and sends this to the signer.
\item
  \label{item-proof}
  The user proves to the signer that she knows $s$ and $r_u$ used to construct
  $\id{blind}(s,r_u)$ using a three messages zero-knowledge protocol. The signing protocol aborts if this proof fails.
\item
  \label{item-sign}
  The signer generates some randomness $r_s$ and creates an intermediate signature $i = \id{intermediate}(b,p,r_s,\privkey{})$ using its private key $\privkey{}$ over the blinded information $b$ received from the user, the
  public part $p$ of the to be signed message, and the randomness $r_s$ it just generated. The signer sends this intermediate signature to the user.
\item
  \label{item-transform}
  The user transforms this intermediate signature $i$ to the final partially blind signature $\blindsig{s}{p}{\privkey{}}$. The user acknowledges this to the signer
\end{enumerate}
Both the user and the signer keep a record of the values of all local variables used and messages exchanged during the signing protocol, and keep track of when they aborted the protocol. Current values of local variables must be safely stored before sending any message that depends on them. If both parties successfully complete the protocol, both can destroy the record for the protocol.

Observe that the only dispute that can occur is when a user claims not to have received a blind signature in return for a payment or a receipt.\footnote{%
  This uses the fact that the signer is honest. The idea is that if the signer claims not to have received the payment or the receipt, then any clearing and settlement of the payment or use of such a receipt will be detected later, and would lead to legal measures.
  }
Then the following cases have to be considered.
\begin{itemize}
\item
  If the signer aborts before sending $i = \id{intermediate}(b,p,r_s,\privkey{})$ (the intermediate signature), then the protocol can be restarted from scratch. This results in a different blind signature, possibly for a different blind secret input $s$, but the same public input $p$. But since it is guaranteed that the sender never sent the intermediate signature,
  we are certain the user was never able to obtain a blind signature in the aborted run.
\item
  If the signer aborted after generating the intermediate signature $i = \id{intermediate}(b,p,r_s,\privkey{})$ (and this intermediate signature may or may not have been received by the user), then the protocol must be picked up from this point, with the user using the stored values for the variables used in step~\ref{item-blind} and \ref{item-proof} (which should exist by assumption that local variables must be safely stored before sending messages that depend on them. This means that the previously generated intermediate signature $\id{intermediate}(b,p,r_s,\privkey{})$ is sent to the user. This results in possibly a different blind signature, but for the same blind secret input $s$ and same public input $p$ that were used in the aborted run.
\end{itemize}
We conclude that the above sketched dispute resolution protocol allows the user to obtain a valid blind signature of her choice (if the dispute resolution protocol itself does not abort of course), while guaranteeing that a (dishonest) user is never able to obtain two different blind signatures for two different values $s$ and $s'$.

\subsection{Attribute based credentials}
\label{ssec-abc}

Partially blind signatures allow the user to hide (part of) the contents of a message to be signed, but must always reveal the full contents of the signed message to allow the signature to be verified. This means that such signatures \emph{only} break the link between the signing and the verification of the messages, meaning that the act of signing and the act of verifying is unlinkable. Unfortunately, any two acts of verification can still be linked (using the unique data embedded in each signature). 

For so-called multi-show unlinkability full blown attributed based credentials are required~\autocite{idemix2012,camenisch2001anoncred}. We will not go into the details here, but only describe the functionality offered by such credentials, and the privacy properties they entertain. Such attributed based credentials are used in the protocols below to implement travel passes and seasonal tickets that offer reduced fares and that, by their very nature, are on the one hand tied to a particular person while on the other hand need to be presented continually to claim a reduced fare.

An attribute based credential is a secure container for one or more attributes $\attrs$. Credentials are bound to a particular person, and the attribute(s) it contains describe certain properties of that person. (In the current context, it describes the eligibility to certain fare reductions, for example because the person is more than 65 years old, or because the person is a student.) The values for the attributes are negotiated by the requesting person and the \term{issuer} $\issuer$ (under the assumption that the issuer knows or can verify that a particular property holds for the person to which the credential is being issued). The issuer also signs the credential, to prevent fraud. We write $\cred{\attrs}{\issuer}$ for the resulting credential that the person obtains. Typically the credential also contains a hidden private $\privkey{\user}$ key known only to the user that is hidden from the issuer when the credential is being issued, somewhat similar to how partially blind signatures work. Tying this private key to the credential and requiring its use when showing the credential later (see below) aims to prevent users from sharing their credentials to commit fraud (\eg when a student allows her younger, non student, brother to use her credential to obtain a reduced fare ticket). We note that such techniques to bind people to their credentials are not fool proof~\autocite{camenisch2001anoncred}.

To prove a certain attribute, the user engages in a so called interactive
\term{showing} protocol with a verifier using one or more of such credentials. This showing protocol is typically \emph{selective}: the user can decide which attributes to reveal to (and which ones to hide from) the verifier. This means that the verifier never gets to see the full credential, which would be a bad idea anyway as every credential signature is unique and therefore would allow subsequent uses of the same credential by the same user to be linked. As we want multi show unlinkability, the user and the verifier instead engage in an (interactive) zero knowledge protocol where the user proves to the verifier that she owns a credential signed by a certain issuer, containing a selection of the revealed attributes $A_r \subseteq \{\attrs\}$. This proof also requires the user to know the embedded private key $\privkey{\user}$ (without revealing it of course). This reveals the issuer and the attribute values, and nothing more, to the verifier.
We write $\privkey{\user},\cred{\attrs}{\issuer} \leftrightarrow \issuer,A_r \subseteq \{\attrs\}$ (where the left hand side shows the input of the user, and the right hand side shows what the verifier learns (provided it knows the public key of the issuer needed to verify the proof).

\subsection{Privacy friendly payment with receipt}
\label{ssec-pay}

A basic mechanism used throughout our protocols is the possibility to pay a
certain fare $\fare$ to a payment service provider (PSP) and to receive a
receipt $\receipt$ for this payment in return.\footnote{%
  The PSP could be your bank (provided it knows how to issue receipts as explained below), or a separate entity that lets bank process the payment and generates a receipt when the payment was successful.
}
The receipt can subsequently be used at a (public transport) service provider to pay for transport. The idea is that such a payment mechanism can be implemented in many different (more or less privacy friendly) ways, with only the receipt being standardised for use in the protocols below. 

To maximise privacy protection in case the payment itself is less privacy friendly, the receipt $\receipt = \receiptdetails$ is a blind signature over the public fare $\fare$ paid as well as a blind receipt sequence number $\receiptseqno$ provided by the user, signed by the payment service provider PSP that processed the payment. To make explicit at which particular service the receipt can be used, the name of the service can be added, blindly, by the user as well. The user should ensure that each receipt has a different sequence number. This sequence number is used to prevent reuse of receipts: the sequence numbers in redeemed receipts are recorded as spent. (Which also shows why users have every reason to ensure that sequence numbers are indeed different.)

Using a blind signature in this way guarantees that users cannot create fake receipts, while the receipt sequence number cannot be linked to the payment (and hence to the user making the payment). Users are expected to properly protect their receipts and keep them securely stored until use.

In the protocols below, the paid fare is first collected by the PSP, then forwarded to a public transport clearinghouse (PTC) that later redistributes the paid fares to the $\PTO$s based on submitted receipts the PTOs have collected. Each fare is recorded by the payment service provider (PSP) as a separate payment transaction for the specified amount with the clearinghouse as the recipient.  If the payment transaction involves the bank account of the user (see below), care should be taken to \emph{not} include the bank account details of the user in the transaction towards the clearinghouse. This happens more or less automatically if the PSP is a separate entity independent of the bank (in which case the transaction will transfer the fare amount from the user bank account to that of the PSP). If the bank itself serves as PSP, an internal bank offset account should be used that aggregates individual payments to the PTC with only the daily or weekly totals being transferred to the actual PTC account. This prevents the clearinghouse from learning the bank account (and hence the identity) of all people travelling with public transport, including how often they travel and an indication of the distance they travel (given that the fare is often a good indication of this).

One possible way to implement payment when using a smartphone based public transport ticketing app is to redirect the payment phase to a separate payment app on the user's smartphone, and let PSP forward the resulting receipt back to the transport ticketing app. A more privacy friendly option is to allow travellers to pay with cash at designated kiosks at public transport stations. Or to support the payment of fares using some kind of online privacy friendly payment scheme (like Digicash~\autocite{DBLP:conf/crypto/ChaumFN88}, or Zcash~\autocite{DBLP:journals/iacr/Ben-SassonCG0MTV14}).

\subsection{Notation}

When describing the knowledge acquired by parties involved in the (figures depicting the) protocols below, we use expressions like $(a,b,c)$ to denote that a party learns the values $a$, $b$, and $c$, and moreover learns that they are linked and thus belong together. Values in different tuples are not linked, but can however be correlated based on their actual values: if a party learns a specific fare $\fare$ was paid by user $\user$ (\ie it knows $(\user,\fare)$)
and later sees a ticket with that particular fare for a trip $\trip$ (\ie it also knows $(\tripdetails,\fare)$, then it may conclude user $\user$ travelled route $\triproute$ on date $\tripdate$. We use $\useraddr$ to denote the possibly static IP network address of the user visible to the other parties.\footnote{%
  This equals the VPN server address or the Tor exit node address in case any of these services are used by the user.
}

\begin{figure*}[t]
  \centering
  \setoverpicfontsize
  \begin{overpic}[abs,unit=1pt]{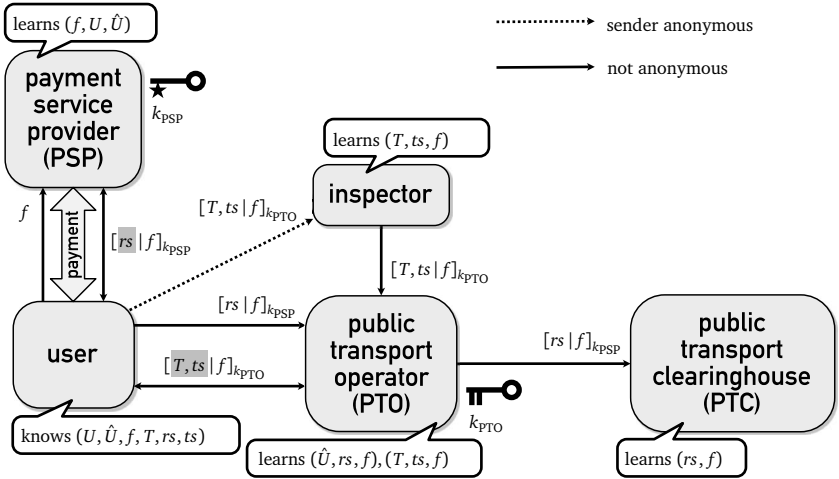}% 
  \def\f{$\fare$}
  \def\r{$\receiptdetails$}  
  \def\rr{\llap{$\usereceiptdetails$}}
  \def\tt{\llap{$\useticketdetails$}}
  \def\t{$\ticketdetails$}  
  \def\ti{$\useticketdetails$}
  \def\kb{$\privkey{\PSP}$}
  \def\kpto{$\privkey{\PTO}$}
  \def\bl{learns $(\fare,\user,\useraddr)$}
  \def\ul{knows $(\user,\useraddr,\fare,\trip,\receiptseqno,\ticketseqno)$}
  \def\il{learns $(\trip,\ticketseqno,\fare)$}
  \def\pl{learns $(\useraddr,\receiptseqno,\fare),(\trip,\ticketseqno,\fare)$}
  \def\cl{learns $(\receiptseqno,\fare)$}
  \def\la{sender anonymous}
  \def\lb{not anonymous}
  \input{./fig/prot-papertickets.overpic}%
  \end{overpic}
  \caption{Protocol emulating paper tickets}
  \label{fig-prot-papertickets}
\end{figure*}

\section{Solution 1: Emulating paper tickets}
\label{sec-papertickets}

One way to achieve privacy in public transport ticketing is to emulate the traditional use of paper tickets in public transport. The basic idea is to first buy the ticket online, and subsequently use it for public transport later, in such a way that the financial transaction used to pay for the ticket cannot be linked to the actual trip being made. The protocol assumes that the public transport app on the user's smartphone contains a database with all possible trips that can be made by public transportation, together with the corresponding fares to be paid.

\subsection{Detailed protocol}

The protocol, graphically represented in figure~\ref{fig-prot-papertickets}, runs as follows.

\begin{description}
\item[Phase 1: Obtaining a ticket]~
  
  \begin{itemize}
  \item
    The user selects the route $\triproute$ she wants to travel, and the day $\tripdate$ on which she wishes to travel. This defines the trip
    $\trip = \tripdetails$.
  \item
    The user calculates the fare $\fare = \farefor{\trip}$ for the trip. (Incorrectly calculated fares will be detected later.)
  \item
    The user starts a payment for this fare, and receives a receipt $\receipt = \receiptdetails$ in return. (See section~\ref{ssec-pay} above for details.)
    The paid fare is credited to the PTC account.
   \item
    The user sends this receipt to the PTO. The PTO verifies the signature on
    the receipt, and submits it for clearing and settlement to the PTC. The PTC
    also checks the signature on the receipt, and checks whether a receipt with
    sequence number $\receiptseqno$ has been submitted before. If so, the
    receipt is rejected. Otherwise, the PTC accepts the receipt and records
    $\receiptseqno$ as submitted.
  \item
    The user engages in a partially blind signature issuing protocol with the PTO in order to obtain a ticket $\ticket$ for the trip. The user blindly provides the trip $\trip$ as well as a blind and fresh ticket sequence number $\ticketseqno$. The PTO provides the (unblinded) fare $\fare$ present in the receipt it received in the previous step. As a result the user receives the ticket $\ticket = \ticketdetails$, signed by the PTO.
  \end{itemize}

\item[Phase 2: Travelling by public transport]
  Public transport operators need to verify that all users that travel with
  them have a valid ticket, with the correct fare. The traveller and the ticket inspector engage in the following protocol for that purpose.
  \begin{itemize}
  \item
    The user sends the ticket $\ticket = \useticketdetails$ (revealing all its
    contents) to the ticket inspector. One way to do so in a sender anonymous fashion is to let the public transport app display the ticket as a QR code on the smartphone display, and let the inspector scan this QR code. Many public transport operators use similar schemes to inspect `home print' paper based tickets.
  \item
    The ticket inspector verifies the signature on the ticket, whether fare
    $\fare$ is correct for trip $\trip = \tripdetails$, whether the ticket
    sequence number $\ticketseqno$ is not invalidated, whether the date
    $\tripdate$ in trip $\trip$ is today, and whether the route $\triproute$ in
    trip $\trip$ covers the leg (of the total trip) where the ticket inspector
    asks the user to provide a ticket. If the user cannot provide a valid
    ticket, a fine is issued.
  \item
    The ticket inspector verifies the ticket with the PTO. The PTO
    also checks the signature on the ticket, and checks whether a ticket with
    sequence number $\ticketseqno$ has been submitted before. If so, the
    ticket is rejected. Otherwise, the PTO accepts the ticket and records
    $\ticketseqno$ as submitted. (If trips consist of several legs, the same
    ticket should be accepted for different legs of the trip.)
  \end{itemize}
  Note that the fact that ticket sequence numbers must be verified and
  invalidated in real time implies that the equipment of the inspector must be
  online. As tickets are only valid for a single day, PTOs may choose to forfeit on this strict form of checking (hence relaxing system requirements), relying on the fact that ticket can still only be used (perhaps multiple times) on a single day.
  
\item[Phase 3: Clearing and settlement]
  PTOs are reimbursed based on the payment receipts received in phase 1, after
  submitting them to the clearinghouse PTC. For each receipt, the PTC verifies
  the signature, and verifies that the sequence number in the receipt
  $\receiptseqno$ is yet unclaimed. If so, the sequence number is recorded as
  claimed, and the PTC proceeds to pay the fare specified in the receipt to the
  PTO. Otherwise the claim is rejected.

\end{description}

\subsection{Analysis}
\label{ssec-analysis}

To what extent does this solution fit the requirements set out above?

Users obviously pay have to for their trips, and the fare depends on the distance travelled. Inspectors and sufficiently high fines are necessary to keep users honest and disincentivise travelling without a valid ticket.

Public transport operators get paid based on the payment receipts they collect when issuing tickets. To get (statistical) information about actual trips made they need to have enough conductors to check the tickets of all their passengers when travelling (as this is the only time when the actual trip details are revealed). If multiple PTOs are involved in a particular trip, proper reimbursement can only be achieved if the app splits up the trip in different legs, one for each PTO the user needs to travel with.

The level of privacy protection is reasonable, depending on the properties of the network being used. The protocol prevents trip details to be linked to users, in the following sense: for all the tickets a particular PTO sells for a particular fare $\fare$ it learns the set of IP addresses of users that bought a ticket for this particular fare on the one hand, and the set of trips made for this fare (through inspection) on the other hand, but it can never link a particular user address to a particular trip.

In the setup described, the PSP and PTOs could however learn how many tickets you buy, and for which amount (ie for which distance), if they would try to identify you based on your (fixed) IP address. Note that this problem becomes much less significant if the payment receipts issued by the PSP can be used for many different types of purchases, \ie if they are used as a type of generic digital currency. Even if PTOs and PSP collude, they would not be able to link users with actual trips made, but timing analysis linking payment times with ticket issuing times could be used by the PTO to be more certain about your identity. If you usually buy your tickets on the same day or the day before your trip, your PTO could learn when you travel. The PTO could learn whether you are using public transport a lot, or not. Many short trips on the same day may reveal you are in a city; certain patterns of distances may correspond to popular tourist routes (and hence reveal the city you are in). This limited level of privacy protection may already be a threat for people that engage in protests or civil disobedience, like the Hong Kong protesters or the Extinction Rebellion activists. All these problems can be avoided if users can use cash to buy tickets, at special digital kiosks.

Communication between the user and inspector is sender anonymous. This means ticket inspection reveals no personal information.

The system is secure: tickets are only issued by the PTO when given a payment receipt for a certain amount, signed by a bank. Only banks can create such a signed proof of payment. The amount paid for a ticket is checked by the conductor when inspecting a ticket. Only PTOs can create a valid ticket (signed in partially blind fashion). This signature is also checked by the conductor. Finally the conductor checks whether the ticket entitles a person to travel when and where the conductor inspected her ticket. Failure of one of these tests means the ticket is invalid. The sequence number of the ticket (embedded to guarantee one-time use) is checked in real-time with an online database of sequence numbers of already inspected tickets. If the sequence number is already in the database, the ticket is invalid. Otherwise, the sequence number is added to the database.

\subsection{Dealing with failures and disputes}
\label{ssec-prot1-failures}

Dealing with failures is always a challenge, but this is particularly the case in privacy friendly protocols where often the link between a user and her actions is deliberately broken. This means extra care needs to be taken to create some evidence that allows an entity to challenge a failure, while not eroding the privacy of the users. Below we describe some possible failures, and how they could be dealt with. See also~\autocite{stubblebine1999unlinkable-transactions} for additional measures that can be taken, and the general strategy to deal with failures during the issuing of blind signatures (like users not receiving a payment receipt after payment, or not receiving a ticket after submitting a payment receipt) outlined in section~\ref{ssec-failures}.

\begin{description}
\item[The user wants to cancel a payment]
  The user can return the payment receipt (which contains a unique sequence number) to the PSP to rewind the transaction. The PSP then forwards the payment receipt to the PTC signalling not to accept this payment receipt when a user requests a ticket to be issued. Also, the transfer of money from the bank to the PTC will be reversed.

\item[The user receives a valid but incorrect ticket]
  This can happen if the user entered the wrong trip details, or if some internal error caused the wrong ticket to be issued. The user can 'return' the ticket to the PTO, essentially running the showing protocol normally run when a conductor inspects the ticket. This invalidates the ticket. Using the same payment receipt she can start restart the issuing step, now with the correct trip details (assuming the fare is the same).

\item[The user wishes to cancel a ticket issued to her]
  After 'returning' the ticket the PTO as described in the previous case, she can then proceed to cancel the payment to the bank.

\item[The user receives an invalid ticket]
  This is more tricky. Ideally the issuing protocol should guarantee that a valid ticket is issued. If this is impossible, at least the issuer should somehow be able to tell, from the logs, that the user indeed did not receive a valid ticket. Otherwise bogus claims for invalid tickets could be submitted. This all very much depends on the particular issuing protocol used.

\item[A valid ticket fails conductor inspection]
  Ideally this should not happen. However, the user or conductor device may malfunction, and the communication between the two devices may be erroneous. If the ticket is valid, and the user app operates correctly, the user should at some point be able to convince the PTO she had a valid ticket when travelling. 

\item[The app crashes or malfunctions]
  This can be mitigated by ensuring that the app can be reinstalled without loosing any stored tickets, or turning them invalid. This requires operating system support, \eg allowing data to be restored from data associated with a previous install of the application.

\item[The user looses or deletes a ticket]
  There is no way to recover from this situation. (Loosing a ticket could happen when inadvertently deleting the whole app together with all its data.) 
\end{description}

\subsection{Variations and extensions}

\subsubsection{Using actual paper tickets, or smart cards}

Instead of relying on users having smartphones, tickets could actually be
printed on paper,\footnote{%
  This may sound pedantic, but in fact when trying to emulate
  something digitally based on how it was done physically, one always has to consider the option that the original, physical, approach simply works better.
}
or be stored on contactless smart cards instead. In this case, a ticket kiosk needs to be used to allow users to select the ticket they need, allow them to pay (by cash or card), and to print the ticket or issue the ticket to the smart card. In the first case, the ticket (with its signature) is printed as QR code, which the inspectors can scan with their smartphone. In the second case, inspectors need to carry NFC enabled smartphones that allow them to scan the smart card and read the ticket (with its signature) from the smart card. This is certainly possible even with cheap smart cards (as it is not involved in any complex cryptographic operation: the inspector checks the signature locally on the device, and the blind signature is generated by the kiosk where the user buys the ticket). Paper tickets can also be obtained at home through a website (web app) that essentially emulates the functionality of the user smartphone app with respect to obtaining a ticket, but at the end of this phase prints the ticket as a QR code instead of storing it.

\subsubsection{Supporting seasonal tickets}

Reduced fares for public transport pass subscribers or holders of seasonal tickets can be catered for in a privacy friendly manners using attribute based credentials, in which case the attributes in the credential encode the fare reductions the holder is entitled to. The user can obtain such a credential using a protocol similar to that of buying a single ticket, except that in the last step the PTO issues a full blown credential instead.\footnote{%
  A simple blind signature as used for ordinary tickets will not do as the credential will have to be shown multiple times while retaining the desired privacy properties.
}

The issuing protocol would run like this.
\begin{itemize}
\item
  The user selects which type of seasonal ticket she wishes to buy. This defines a set of attributes $\attrs$ that define which type of reduction she is entitled to.
\item
  The user calculates the total price $\fare$ for this seasonal ticket.
\item
  The user starts a payment for this amount, and receives a receipt $\receipt = \receiptdetails$ in return. (Again see section~\ref{ssec-pay} above for details.) The paid amount is credited to the PTC account.
\item
  The user sends this receipt to the PTO. The user also sends the list of attributes $\attrs$ to the PTO. The PTO verifies that the price $\fare$ present in the receipt corresponds to the amount due for this particular set of attributes. The PTO verifies the signature on the receipt, and submits it for clearing and settlement to the PTC. The PTC also checks the signature on the receipt, and checks whether a receipt with sequence number $\receiptseqno$ has been submitted before. If some of these tests fail, the receipt is rejected. Otherwise, the PTC accepts the receipt and records $\receiptseqno$ as submitted.
\item
  The user and the PTO engage in a credential issuing protocol for this set of attributes. As a result the user receives the credential $\credentialdetails$, signed by the PTO.
\end{itemize}

Such a credential can subsequently be used to travel by public transport with a reduced fare. Interestingly enough, the credential is actually irrelevant when buying a ticket (except that the user needs to apply the correct fare reduction based on the particular credential she owns), because the PTO blindly issues a ticket for a particular fare without learning the actual trip the ticket is for. Correctness of the fare paid is only verified at inspection time, when the inspector gets to see the full ticket containing both the trip and the corresponding fare. To prove that the user is entitled to a reduced fare, the inspection protocol needs to incorporate verification of the necessary attributes in the credential as well. As the original inspection protocol is sender anonymous, the credential verification protocol needs to be sender anonymous as well. This can be achieved by using a fully non-interactive credential showing protocol. Idemix~\autocite{idemix2012}, for example, uses a non-interactive proof of knowledge, but relies on a verifier generated nonce to guarantee freshness of the proof. Such a verifier generated nonce would be hard to incorporate in our setting, as it would require the equipment of the inspector to send something to the user device (which would either break sender anonymity or would require cumbersome approaches where the user needs to also scan a QR code on the inspector device). Luckily there is a way out of this dilemma: we can use the cryptographic hash of the randomly chosen ticket sequence number $\ticketseqno$ already present in the ticket as the nonce instead. The fact that in this case the nonce is generated by the prover is not a problem but actually a feature: the proof is now neatly tied to the ticket for which a reduced fare is claimed, and the original ticket inspection protocol already ensures that the same ticket sequence number cannot be used twice. This forces the user to pick a fresh sequence number.\footnote{%
  Note that the user is by no means forced to select the ticker sequence number \emph{randomly}. Hashing it to derive the actual nonce to be used in the credential showing protocol however ensures that the protocol remains secure when the underlying credential showing protocol relies on actual randomness (and not merely freshness) of the nonce.
}

A problem with the approach outlined above is that there is nothing inherently preventing users to pool and share a single credential (offering reduced fares) with a group of users that each 'prove' possession of the credential to the inspector when necessary. Unless the private key associated with the credential is securely embedded in the user device (using \eg a piece of trusted hardware to ensure that even the device owner cannot get access to it), this by itself does prevent such credential pooling attacks. This is a general problem of attribute based credentials, and indeed a problem of online digital identity management in general as securely binding actual persons to their online credentials is hard~\autocite{camenisch2001anoncred}.\footnote{%
  Even embedding the private key in a secure enclave does not strictly speaking prevent the owner of the smartphone to share the phone itself with others (although it is surely not an enticing proposition to be without your private phone for several hours).
}

\section{Solution 2: Pay as you go, with credit on device}
\label{sec-pay-as-you-go}

A fundamentally different, and increasingly popular approach for letting people pay for public transportation is to store credit on a contactless smart card serving as a public transport pass. People can (re)charge their passes at special kiosk (essentially transferring money from their bank account to their public transport pass) and subsequently pay when entering or leaving their chosen mode of transportation. This typically involves `checking in' at a gate or turnstile when entering the station, or on the platform or in the bus itself, and `checking out' when arriving at the destination or when changing connections. When people check-in, a check is performed to see whether there is enough credit left on the card. If so the location of the check-in is recorded on the card. When checking out, this check-in location is retrieved, and based on the check-out location the fare is computed and deducted from the credit on the card. To detect fare dodgers that travel without checking in, inspection on the trains or the bus is often still necessary, because it is hard to enforce an air-tight system that forces people to check-in or check-out at all times. The main challenge in implementing such a scheme is to ensure that the check-out operation is performed as fast and reliably as possible (given that at busy transportation hubs many people have to the check-out at the same time, and that a transaction involving a contactless public transport pass is prone to interference and failures).

If current public transportation pass systems would actually work as just described, there would be no need to study privacy friendly forms of public transport ticketing: if all that the cards contain is user credit, there would not by any privacy issues with such a system. Unfortunately, this is not the case. All systems mentioned above involve cards with unique serial numbers that are recorded when checking in and when checking out, and stored in a central database. As these serial numbers are static, this allows users to be singled out and their public transportation travel patterns to be recorded over the years. What's worse: these passes are almost always bound to a particular user (either because they are tied to a personal public transport account, or simply because they were recharged using the bank account of the user). The main reason for adding such tracing of passes is to be able to detect fraud and block passes that appear to be spending more credit than they should be spending based on the amounts used to charge them. 

Here we aim to emulate such a credit-based system in a privacy friendly manner, without needing to rely on tamper proof hardware or secure execution environments to prevent users from committing fraud by tampering with the credit on their tokens (\ie their smartphones) in their possession.

\subsection{Detailed protocol}

\begin{figure*}[t]
  \def\uncenter{\leftskip0pt\rightskip0pt\parindent0pt}
  \setoverpicfontsize
  \centering % puts everything in a vbox centered
  %  \begin{overpic}[abs,unit=1pt]{./fig/prot-PAYG-v3-x.eps}%
  \begin{overpic}[abs,unit=1pt]{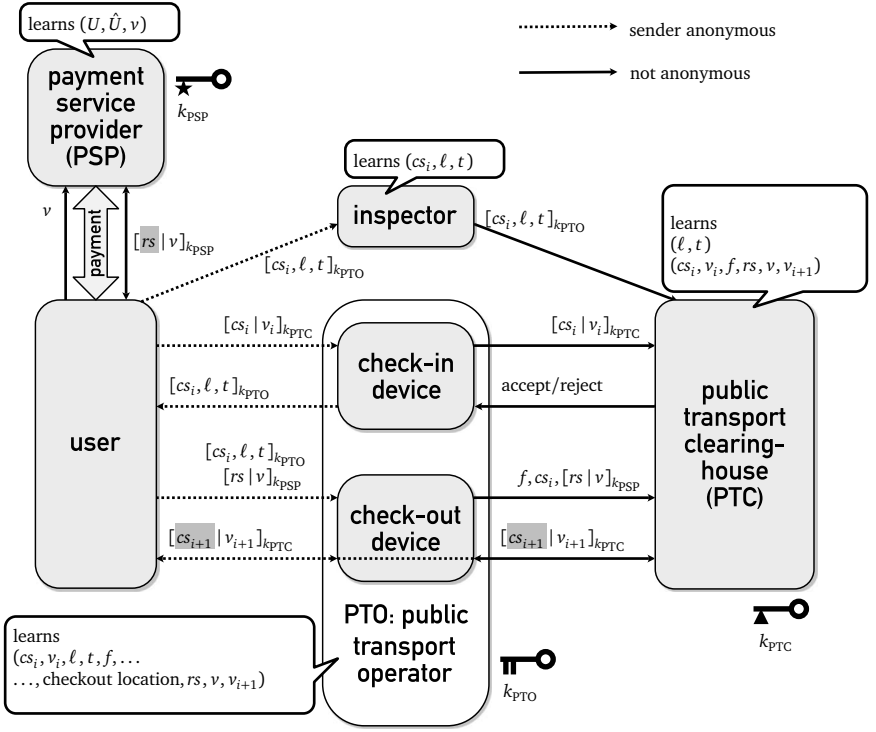}%  
  \def\f{$\creditvalue$}
  \def\r{$\creditreceiptdetails$}
  \def\c{$\creditdetails[i]$}
  \def\i{$\checkindetails$}  
  \def\iptc{$\checkindetails$}
  \def\ci{\llap{$\usecreditdetails[i]$}}
  \def\ciptc{\llap{$\usecreditdetails[i]$}}    
  \def\cirptc{accept/reject}
  \def\cir{$\checkindetails$}
  \def\co{\vbox{\uncenter
      $\usecheckindetails$\\
      $\quad\usecreditreceiptdetails$}}
  \def\coptc{\llap{$\fare,\creditseqno[i],\usecreditreceiptdetails$}}
  \def\corptc{$\blindsig{\creditseqno[i+1]}{\creditvalue[i+1]}{\privkey{\PTC}}$}
  \def\cor{$\blindsig{\creditseqno[i+1]}{\creditvalue[i+1]}{\privkey{\PTC}}$}
  \def\kb{$\privkey{\PSP}$}
  \def\kpto{$\privkey{\PTO}$}
  \def\kptc{$\privkey{\PTC}$}
  \def\bl{learns $(\user,\useraddr,\creditvalue)$}
  \def\ul{knows $(\user,\useraddr,\creditvalue,\receiptseqno)$}
  \def\il{learns $(\creditseqno[i],\location,\ctime)$}
  \def\cl{\vbox{\uncenter
      learns\\
      $(\location,\ctime)$ \\ $(\creditseqno[i],\creditvalue[i],\fare,\receiptseqno,\creditvalue,\creditvalue[i+1])$
    }}
  \def\pl{\vtop{\uncenter
      learns\\
      $(\creditseqno[i],\creditvalue[i],\location,\ctime,\fare,\ldots$\\
      $\ldots,\text{checkout location},\receiptseqno,\creditvalue,\creditvalue[i+1])$
    }}
  \def\la{sender anonymous}
  \def\lb{not anonymous}
  \input{./fig/prot-PAYG-v2.overpic}%  
  \end{overpic}
  \caption{Protocol ``pay as you go''}
  \label{fig-prot-PAYG}
\end{figure*}

Each user maintains travel credit on their own device. As the device is not
assumed to be trusted or tamper resistant, care must be taken to ensure that
users cannot create counterfeit credit, or spend more than they have
credit. Therefore, travel credit $\credit = \creditdetails$ is represented by a
blind signature of the Public Transport Clearinghouse $\PTC$ over the secret
(blinded) sequence number $\creditseqno$ and the known credit value
$\creditvalue$. Once the credit token is used, the sequence number $\creditseqno$ becomes known. In the protocol below the PTC uses this to record the 'state' of such a token as either checked-in or spent.

We assume in this protocol that check-in and check-out use a sender anonymous form of communication, for example by using near field communication with a randomised anti-collision identifier. The protocol runs as follows.
\begin{description}
\item[Phase 1: Obtaining credit]~

  \begin{itemize}
  \item
    The user starts a payment for the amount $\creditvalue$ she wishes to
    obtain credit for, and receives a receipt $\receipt = \creditreceiptdetails$ in return. (See section~\ref{ssec-pay} above
    for details.) The paid amount is credited to the PTC account.
  \item
    The user can use this receipt to add the credit to her device when checking out (see phase 3 below).\footnote{%
      A separate protocol between the user device and the PTC to add credit is also possible, but is not discussed here.
    }
  \end{itemize}

\item[Phase 2: Check-in to start a trip]~
  
  \begin{itemize}
  \item
    The user sends her credit token $\credit = \usecreditdetails[i]$ to the check-in device.
  \item
    The check-in device verifies the signature on the credit token, checks that
    the stored value $\creditvalue[i]$ is larger than some minimum credit
    required,\footnote{%
      This is necessary to prevent users to accrue (too much) negative credit by checking in with hardly any credit and going on an expensive trip.
    }
    and submits it to the PTC. The PTC verifies the signature on the credit token, and checks whether $\creditseqno[i]$ is recorded as spent or checked-in. If so, the credit token is rejected. Otherwise it is accepted and the PTC records $\creditseqno[i]$ as checked-in, and records the associated value $\creditvalue[i]$ necessary when issuing a new credit token at check out.
  \item
    The check-in device sends the user a check-in token $\checkin = \checkindetails$, containing the check-in location $\location$, the check-in time $\ctime$ and the credit token sequence number $\creditseqno[i]$, all signed by the PTO. The PTO logs the tokens for bookkeeping purposes. The user stores the check-in token.     
  \end{itemize}

\item[Phase 3: Check out to finish a trip]~

  \begin{itemize}
  \item
    The user sends her check-in token $\checkin = \checkindetails$ to the check-out device.
  \item
    If the user wants to add additional credit to her device, she also submits a receipt $\receipt = \creditreceiptdetails$ obtained earlier.
  \item
    The check-out device verifies the signatures on both tokens, and validates the time on the check-in token (\ie checks whether the check-in time
    $\ctime$ and check-in location $\location$ make sense given the
    check-out time and the check-out location).
  \item
    Given the check-in time $\ctime$, the check-in location $\location$,
    the check-out time and the check-out location, the check-out device computes the fare $f$.
\item
    The check-out device then submits the fare $\fare$, the credit sequence number $\creditseqno[i]$, and the (optional) receipt $\creditreceiptdetails$ to  the PTC. (The PTO signs this transfer). The PTC also verifies the signature on the receipt, and checks whether $\creditseqno[i]$ is recorded as checked-in. If not, the check in is rejected. Otherwise it is accepted and the PTC records $\creditseqno[i]$ as spent. The PTC retrieves the associated credit $\creditvalue[i]$ (stored when the credit token was submitted at check-in) and computes the new credit $\creditvalue[i+1]=\creditvalue[i] + \creditvalue - \fare$. (Negative credit is possible, but controlled through the credit check at check-in.)
\item
    The user engages in a partially blind signature issuing protocol with the
    PTC, using the check-out device as a relay, in order to obtain an updated credit token $\credit[i+1]$. The user provides a blind and fresh credit sequence number $\creditseqno[i+1]$. The PTC provides the (unblinded) credit value $\creditvalue[i+1]$ it just computed. As a result the user receives the new credit token $\credit[i+1] = \creditdetails[i+1]$.
  \item
    The PTC proceeds to pay the fare to the PTO. (This can be done in bulk.)
  \item
    The user stores the new credit token. The user also logs the check-in in a local trip history (that can be consulted to resolve disputes). It may verify locally whether the deducted fare is correct.
  \end{itemize}

\item[Phase 4: Inspection]
  The inspector needs to verify that every person travelling has a valid check-in token.
  \begin{itemize}
  \item
    The user sends her check-in token $\checkindetails$ to the inspector.
  \item
    The inspector verifies the signatures on the token, validates the time on the check-in token (\ie checks whether the check-in time $\ctime$ and check-in location $\location$ make sense given the inspection location,
%    \footnote{%
%      This inspection is less strict than one involving actual tickets, because the only thing known at inspection time are the check-in time and check-in location, and not the final destination. In check-in/check-out based systems used in a nationwide public transport system, like in the Netherlands, people can travel all across the country and travel more or less for free when they check out at a station close to where they checked in. When checking out and checking back in can be avoided at a small far away station, this is an inviting opportunity of fraud. Also the penalty for forgetting to check out (again something that easily happens when checking out is not physically enforced using \eg turnstiles) is a balancing act between preventing fraud by people deliberately not checking out after a long distance trip, and over-fining people that really forgot to check out on a short trip.  
%    }),
    and submits the check-in token to the PTC. The PTC also verifies the signature on the check-in token, and checks whether $\creditseqno[i]$ is recorded as checked-in. If not, the credit token is rejected. Otherwise it is accepted and the PTC records $\creditseqno[i]$ as inspected. (If this particular token is encountered by a different inspector, on a leg of the trip that is inconsistent with earlier inspections of the same token, then fraud is assumed.)
  \end{itemize}
  Note that the fact that credit must be verified in real time implies that the equipment of the inspector must be
  online, communicating with the PTC (and not the PTO).

\item[Phase 6: Clearing and settlement]
  The PTO logs all check-in and credit tokens submitted during check out. The PTC pays the fare as soon as it receives the check-out token and computes the new credit token. (It may accumulate fares to pay the total amount every day or week.) The PTO verifies the payments it receives with the logs it keeps.
  
\end{description}

\subsection{Practical considerations}

As discussed in the introduction of this section, the main challenge in practice of is to make check-in and check-out as fast (and reliable) as possible.

Reliability can be improved in the above protocol by adding an acknowledgement message back from the user to the check-in or check-out device whenever the check-in or check-out token have been received in good order, and letting the check-in or check-out device generate an appropriate sound as confirmation. The user device itself could confirm proper check-in or check-out immediately after receiving the token (and sending the acknowledgement), or sound an alarm when the expected token is not received within a short timeout. But an additional message does increase the time needed to check-in or check-out, and adds another point of failure as well: what to do if the acknowledgement message itself is not delivered?

Check-in speed is constrained both by the real time connection between the user device and the check-in device, and the real time connection between the check-in device and the PTC which needs to verify that the credit token is not double-spent. This check could be made asynchronous, and the check-in token be issues optimistically, at the expense of ramping up inspection within the public transportation system to detect people that checked in with such a double spent credit token. Alternatively, when there are not too many check out devices, optimistically issued check-in tokens can be revoked when necessary by blacklisting the embedded credit sequence number $\creditseqno$ and sending this to all check-out devices. The check-in token itself involves computing a basic signature over the credit sequence number sent by the user device, after verifying the blind signature over the credit token. This should not prove to be an issue in practice.

Check-out is more complex as it involves issuing a blind signature over the new credit, where the check-out device works as a relay between the user device and the PTC. Check-out speed can be significantly improved by decoupling the issuing of the check-out token from updating the credit on the user device, doing it 'lazily' after check-out with a separate protocol that runs between the user device and the PTC. In this case a basic check-out token containing the fare can be issued by the check-out device, with an ordinary signature (instead of a blind one). To protect user privacy however, care needs to be taken to then hide the user address from the PTC to prevent it from linking the previous credit sequence number $\creditseqno[i]$ to this used address (as this allows the full trip to be linked to a particular user).

\subsection{Dealing with failures and disputes}

Beyond the failures and disputes for protocol 1 discussed in section~\ref{ssec-prot1-failures}, the use of of check-in and check-out devices poses additional challenges. Also the fact that credit is stored on the user device makes the solution more fragile and risky for the user.

Dispute resolution depends on clear information about what happened about the time a failure occurred. Unfortunately, due to their privacy friendly nature, the protocols retain very little useful information by themselves. Adding timestamps to local logs of each protocol step, by the PTC, the PTO, and the user device will help compare logs in case of disputes (and detect possible fraud). Creating append only logs (using hash chaining techniques) increases their integrity, especially if occasional public commitments to the current state of the log are recorded. A hash of the log on a user device can be submitted when checking in and checking out, and be included in the check in and check out token (that are signed by the PTO). This poses no linkability as the log will be updated with every check in and check out, provided such updates always contain some private information from the user device (e.g. the serial number used in the next credit token).

To aid dispute resolution, the PTO could also issue a separate check-out $\sig{\creditseqno[i],\location,\ctime,\fare}{\privkey{\PTO}}$ to the user when she checks out, containing the credit sequence number, check-out location, time, and fare, signed by the PTO. This allows the user to verify the correctness of the new credit token she receives when checking out (and allows her to check that the correct check-out information is used to compute the fare). 

\begin{description}
\item[Check in fails]
  If it is a communication error in the first step, the user can try again. Otherwise, if the credit token fails to verify the user needs to start a dispute resolution (if she believes the credit token should be valid). If the credit token is accepted, but subsequent steps fail, dispute resolution should clear the recorded serial number for the credit token from the clearinghouse database to ensure it is valid the next time the user checks in.

\item[Check out fails]
  If it is a communication error, the user can try again. Otherwise, if the check-in token (or payment receipt) fails to verify the user needs to start a dispute resolution (if she believes the check-in token should be valid).  If the check in token is not accepted, dispute resolution needs to determine whether the user actually tried to check in earlier, or did not. If the user did not get an error when checking in, for sure the PTO log will contain the serial number of the current credit token.

\item[Fare dispute]
  After checking out user discovers that the fare paid does not correspond to the fare due for the trip made. The user should submit a piece of the log with all entries involving the check in and corresponding check out for this trip (which should follow each other immediately in the user device log, and are thus linked through the internal hash chain). This is then matched with the corresponding logs of the PTO and clearinghouse. Any discrepancy can be compensated by adding it to the current credit on the device by issuing a new credit token. This can be done even after the user has made other, more recent, trips.
\end{description}

\subsection{Analysis}

The security analysis is similar to that of the previous protocol, as presented in section~\ref{ssec-analysis}. We therefore focus on the privacy aspects here.

A significant improvement over the previous protocol is that neither the PTO nor the PTC obtains information about the user identity: communication with the inspector and the check-in or check-out devices is sender-anonymous. PTOs can link check-in and check-out location (and hence trips) to credit sequence numbers, but PTOs cannot link these to anything else (either on their own, or when colluding with others). Credit sequence numbers are in essence ephemeral identifiers.

The situation changes slightly when a user decides to buy additional credit and to add it when checking out. In that case, the PTC learns also the value $\creditvalue$ of the additional credit which can be linked to a particular user when thew PTC colludes with the PSP and the particular credit bought is more or less unique. This can be mitigated by allowing users only to buy predefined values of credit, thus ensuring a reasonable anonymity set of users all buying the same credit at roughly the same time. (We implicitly assume here that a user buys credit well in advance to prevent timing correlation attacks.)

The situation also changes when the credit values stored in a credit token are unique. This would allow the PTO to link a check-in with credit $\creditvalue[i]$ with a subsequent check-in with credit value $\creditvalue[i+1]$. With a bit of 'luck' a PTO might be able to link several trips made by the same user this way. This chain is severed as soon as a common credit value is reached,\footnote{%
  Under reasonable assumptions this would not be a concern in practice. Suppose the maximum credit is $\euro 100$ and fares are multiples of $10$ cents, then there are $1000$ different possible credit values. If there are one million users, the anonymity set would on average contain $1000$ people (although the distribution is probably skewed with larger anonymity sets for smaller credit values).
  The system could also define some default credit value options (like $\euro 25$, $\euro 50$, and $\euro 100$) and nudge users to always top-up their credit to these defaults. The anonymity sets for these particular values would then be much larger.
}
or when a user decides to travel with a different, non colluding, PTO.

\subsection{Pay as you go, paying later}

Given the potential benefits of `pay as you go', it would especially be nice to allow users to pay for their trips afterwards, instead of forcing them to lock significant funds on the device itself. However, introducing a pay later option creates a risk for PTOs as users may fail to pay their debts, so mitigation strategies need to be considered.

The basic idea is use to the same protocol, but allowing negative credit. The main risk is that users use their device up to the maximum negative credit, then de-install the app from their smartphone, and then reinstall a fresh one with a balance of zero. To counter such sybil-like attacks, reinstalling an app should be hard. One way to do so is to tie the install to your device identity or app store identity. In that case the app provider or even the app store itself could start asking questions when someone repeatedly installs the app. But this is not as straightforward as it seems, because ideally we want to allow arbitrary third parties to provide public transport apps (to increase trust).

One idea is that any (third party) app must be 'blessed', by the clearinghouse, with an 'admission credential'. In other words, a user can install any app he or she desires, but all protocols outlined above first verify whether the user has a valid admission credential. The user can obtain this credential, through the app, by registering the app with the clearinghouse.\footnote{%
  However, there should be a way to tie this credential to the specific device being used, to prevent cloning.
  %See also the discussion in the next section on a way to sidestep this problem.
}
This registration process requires the user to prove his or her identity (for example using a government wide digital identity scheme). Note that relying on such an approach is risky, as it undermines the main message that the public transport app is privacy friendly: if that is supposed to be the case, why does it require me to sign in with a government approved digital identity? 

The admission credential is special, because it can be blacklisted: the clearinghouse keeps information about all credentials it issued so that when a user wants to obtain a new admission credential (because he or she claims to have lost their phone, reinstalled the app or whatever), then the previous admission credential becomes blacklisted. Information about the blacklisted credential is sent to all PTOs so that when they check whether some user has a valid admission credential (in the first step of each protocol), this will fail for all blacklisted credentials. Note however that this will not deteriorate the privacy protection offered by the protocols, at least not for users without blacklisted credentials: for every credential that is \emph{not} blacklisted, the PTOs have no way to trace or link valid admission credentials that are not yet blacklisted. The exact privacy properties depend on the specific method to blacklist credentials: a naive scheme might allow the clearinghouse to share blacklisting information about \emph{all} users to the PTO to make them all traceable. The most privacy friendly scheme doesn't even allow blacklisted users to be linked or identified~\autocite{camenisch2002dynamic-accumulators,tsang2007blacklist-credentials}.
  
\section{Conclusions}
\label{sec-concl}

In this paper we explored options how to implement privacy friendly ticketing for public transport in practice. We show that this certainly possible, with certain constraints (or issues that deserve further study, see below). Starting point is the observation that from a privacy perspective it is better to collect personal data locally on the user device, instead of centrally on the servers of the service providers. 

Two different approaches (buying tickets beforehand, and pay as you go) have been studied. We show that these can be implemented with reasonably good privacy properties, under reasonably practical assumptions. In particular we show that an untrusted smartphone can be used as the 'token' to carry tickets or travel credit. This allows third parties to provide the apps for that purpose, which should increase the (perceived) trust of the overall system. For the second protocol, there are rather strict requirements on the maximum checking in and checking out time (in the order of 200-300 milliseconds); actual implementations of the protocols proposed are necessary to verify whether these requirements can be met.

One meta conclusion of this work is that we need an efficient, frictionless, way to provide sender anonymity on the Internet, similar to the use of randomised MAC addresses on local networks. A VPN is too weak (the VPN provider sees everything its users do), yet Tor is too strong (there is no need to protect against a NSA like adversary) given the impact on performance. If randomised client IP addresses could be used by default to set up a TCP connection between a client and a server, that would already provide a tremendous boost in privacy on the Internet as servers can no longer trace their users based on their IP address. There are some proposals for temporary IPv6 addresses that partially address this issue~\autocite{RFC4941bis}, but these only apply to larger subnets and do nothing to hide the often fixed IP addresses of private xDSL connections.

The second meta conclusion of this work is that there is a need to make apps (or data in apps) \emph{uncloneable}, so that they can be used in similar contexts and with similar properties as smart cards. Moreover, there should be a secure way to establish that the person holding the phone and/or using the app is indeed the owner of the phone (and not someone that uses the phone with permission of the real owner). These properties are also mandatory to increase the security of attribute credentials, in particular to prevent the attributes in them being pooled or shared. This seems challenging if at the same time we want the apps to be open source. One idea is to use either the SIM card present in most smartphones, or to use the secure element present in most modern smartphones.

% I'd like to thank Hanna Schraffenberger and Sietse Ringers for useful comments and suggestions on earlier drafts of this paper.

%
% use Makefile.main and friends to extract bib entries from main bib files
% and create a local .bib file; add its name here
%
% strings do not need to be included here (better portability); they are 
% merged into the local bibfile by Makefile.jhh
%
\sloppy
\printbibliography
\end{document}